\documentclass[preprint,tightenlines,showpacs,showkeys,floatfix,nofootinbib,superscriptaddress,fleqn]{revtex4}
\usepackage{graphicx}
\usepackage{bm}
\usepackage{dcolumn}
\usepackage{amsmath}%
\usepackage{amsfonts}%
\usepackage{amssymb}

\begin{document}

\preprint{PNU-NTG-08/2004}
\preprint{TPJU-10/2004}%
\title{~~\\~~\\Octet, decuplet and antidecuplet magnetic moments in
the chiral quark soliton model revisited} 
\author{Ghil-Seok Yang}
\email{gsyang@pusan.ac.kr}
\affiliation{Department of Physics, and Nuclear Physics \&
Radiation Technology Institute (NuRI), Pusan National University, 609-735
Busan, Republic of Korea}
\author{Hyun-Chul Kim}
\email{hchkim@pusan.ac.kr}
\affiliation{Department of Physics, and Nuclear Physics \&
Radiation Technology Institute (NuRI), Pusan National University, 609-735
Busan, Republic of Korea}
\author{Micha{\l} Prasza{\l}owicz} 
\email{michal@th.if.uj.edu.pl}
\affiliation{M. Smoluchowski Institute of Physics, Jagellonian University, ul.
Reymonta 4, 30-059 Krak{\'o}w, Poland}
\author{Klaus Goeke}
\email{Klaus.Goeke@tp2.ruhr-uni-bochum.de}
\affiliation{Institut f\"ur Theoretische Physik II, Ruhr-Universit\" at Bochum,
D--44780 Bochum, Germany}

\date{October 2004}

\begin{abstract}
~~~\newline We reanalyse the magnetic moments of the baryon octet,
decuplet, and antidecuplet within the framework of the chiral
quark-soliton model, with SU(3) symmetry breaking taken into
account. We consider the contributions of the mixing of higher
representations to the magnetic moment operator arising from the
SU(3) symmetry breaking. Dynamical parameters of the model are
fixed by experimental data for the magnetic moments of the baryon
octet and from the masses of the octet, decuplet and of
$\Theta^{+}$. The magnetic moment of $\Theta^{+}$ depends rather
strongly on the pion-nucleon sigma term and reads $-1.19\,{\rm
n.m.}$ to $-0.33\,{\rm
  n.m.}$ for $\Sigma_{\pi N} = 45$ and $75$ MeV respectively. The
recently reported mass of $\Xi^{--}_{\overline{10}}(1862)$ is
compatible with $\Sigma_{\pi N} = 73$ MeV. As a byproduct the strange
magnetic moment of the nucleon is obtained with a value of
$\mu^{(s)}_N =+0.39$ n.m.

\end{abstract}
\pacs{12.39.Fe,13.40.Em,12.40.-y, 14.20.Dh\\
Key words: Pentaquark baryons, magnetic moments, nucleon strange
magnetic moment, chiral soliton model} \maketitle

\section{Introduction}

A recent discovery of the exotic pentaquark $\Theta^{+}$ state
($\mathrm{uudd}\bar {\mathrm{s}}$)  by the LEPS
collaboration~\cite{Nakano:2003qx} and its further confirmation by
a number of other experiments \cite{experiments}, together with an
observation of exotic $\Xi_{\overline {10}}$ states by the NA49
experiment at CERN~\cite{Alt:2003vb}, though it is still under
debate, opened somewhat unexpectedly a new chapter in baryon
spectroscopy. Experimental searches for these new states were
motivated by the theoretical prediction of the chiral
quark-soliton model~\cite{Diakonov:1997mm}, where masses and decay
widths of exotic antidecuplet baryons were predicted. In fact,
exotic SU(3) representations containing exotic baryonic states are
naturally accommodated within the chiral soliton
models~\cite{Chemtob,manohar,Praszalowicz:2003ik}, where the
quantization condition emerging from the Wess-Zumino-Witten term
selects SU(3) representations of triality zero~\cite{su3quant}.

The findings of the pentaquark baryon $\Theta^{+}$ and possibly of
$\Xi_{\overline{10}}$ have triggered intensive theoretical
investigations which are summarized in
Refs.\cite{Jennings:2003wz,Zhu:2004xa}. In particular the
production mechanism of the $\Theta^{+}$ has been discussed in
Refs.\cite{Ohetal,Nametal,Ko:2003xx,Yu:2003eq}. It is of great
interest to understand the photoproduction of the $\Theta^{+}$
theoretically, since the LEPS and CLAS collaborations used photons
as a probe to measure the $\Theta^{+}$. In order to describe the
mechanism of the pentaquark photoproduction, we have to know the
magnetic moment of the $\Theta^{+}$ and its strong coupling
constants. However, information on the static properties such as
antidecuplet magnetic moments and their strong coupling constants
is absent to date, so we need to estimate them theoretically.
Recently, two of the present authors calculated the magnetic
moments of the exotic pentaquarks in a \emph{model-independent
approach}, within the framework of the chiral quark-soliton
model~\cite{Kim:2003ay} in the chiral limit.  
Since we were not able to fix all the parameters for the magnetic
moments in the chiral limit, we had to rely on the explicit model
calculations~\cite{Kim:1997ip,Kim:1998gt}.

The {\em model-independent approach} was introduced 
for the first time by Adkins and Nappi \cite{AdNap} in the context of
the Skyrme model. In this approach, dynamical quantities like moments
of inertia or coefficients in the magnetic moment operator that are
in principle calculable within the model are not numerically
evaluated but treated as free parameters. Adjusting them to the
experimentally known magnetic moments, we allow for maximal
phenomenological input and minimal model dependence. 

The discovery of $\Theta^{+}$ and possibly of
$\Xi_{\overline{10}}$ constrained the parameters of the chiral
quark-soliton model that were previously undetermined. This new
phenomenological input reduces the residual freedom in the
predictions of static baryon properties evaluated in the
\emph{model-independent approach}.

In this paper we revise  
previous results both for nonexotic ~\cite{Kim:1997ip,Kim:1998gt}
and exotic baryons~\cite{Kim:2003ay}.  We show that magnetic
moments of nonexotic baryons (\emph{i.e.} decuplet, since octet
magnetic moments are used as an input) are little changed. On the
contrary, antidecuplet magnetic moments are different from our
previous analysis done in Ref.\cite{Kim:2003ay}.  In particular,
our present study shows that the magnetic moment of $\Theta^{+}$
is negative and rather sensitive to the residual freedom which we
parameterize in terms of the pion-nucleon sigma term: $\Sigma_{\pi
N}$.

The paper is organized as follows. In Sect.~\ref{constraints} we
recapitulate mass formulae within the chiral quark-soliton model
and discuss in some detail the constraints on the model parameters
that come from the measurement of the mass of $\Theta^{+}$ and, if
one wants, of $\Xi_{\overline{10}}$. In Sect.~\ref{magmoms} we
give explicit formulae for the antidecuplet magnetic moments and
display some useful intermediate results in the {\em
model-independent approach}. Numerical results and comparison with
other models are presented in Sect.~\ref{numres}. Finally we
summarize in Sect.~\ref{summary}.
%%%%%%%%%%%%%%%%%%%%%%%%%%%%%%%%%%%%%%%%%%%%
\section{Constraints from the exotic states}
\label{constraints}
%%%%%%%%%%%%%%%%%%%%%%%%%%%%%%%%%%%%%%%%%%%%

The collective Hamiltonian describing baryons in the SU(3) chiral
quark-soliton model takes the following form \cite{Blotz:1992pw}:
\begin{equation}
\hat{H}=\mathcal{M}_{\rm sol}
+\frac{J(J+1)}{2I_{1}}+\frac{C_{2}(\text{SU(3)})-J(J+1)
-\frac{N_{c}^{2}}{12}}{2I_{2}}+\hat{H}^{\prime},  
\label{Eq;1}
\end{equation}
where $\mathcal{M_{\rm sol}}$ and $C_2({\rm SU(3)})$ denote the
classical soliton mass and the SU(3) Casimir operator, respectively.
$I_1$ and $I_2$ are moments of inertia of the soliton.  
The symmetry-breaking term in Eq.(\ref{Eq;1}) is expressed by
\begin{equation}
\hat{H}^{\prime}=\alpha D_{88}^{(8)}+\beta Y+\frac{\gamma}{\sqrt{3}}%
D_{8i}^{(8)}\hat{J}_{i},\label{Hsplit}%
\end{equation}
where parameters $\alpha$, $\beta$ and $\gamma$ are of order
$\mathcal{O}(m_{s})$. Here $D_{ab}^{(\mathcal{R})}(R)$ denote SU(3) Wigner
rotation matrices and $\hat{J}$ is a collective spin operator. The
Hamiltonian given in Eq.(\ref{Hsplit}) acts on the space of baryon wave functions
$\left| \mathcal{R}_{J},B,J_{3}\right\rangle $:
\begin{equation}
\left|  \mathcal{R}_{J},B,J_{3}\right\rangle =\psi_{(\mathcal{R}%
;Y,T,T_{3})(\mathcal{R}^{\ast};-Y^{\prime},J,J_{3})}=\sqrt{\mathrm{dim}%
(\mathcal{R})}(-1)^{J_{3}-Y^{\prime}/2}D_{Y,T,T_{3};Y^{\prime},J,-J_{3}%
}^{(\emph{R})\ast}(R).\label{Eq:wave_f}%
\end{equation}
Here, $\mathcal{R}$ stands for the allowed irreducible representations of the
SU(3) flavor group, \emph{i.e.} $\mathcal{R}=8,10,\overline{10},\cdots$ and
$Y,T,T_{3}$ are the corresponding hypercharge, isospin, and its third
component, respectively. Right hypercharge $Y^{\prime}=1$ is
constrained to be unity for the physical spin states for which $J$ and
$J_{3}$ are spin and its third component.  The {\em model-independent
approach} consists now in using Eqs.~(\ref{Eq;1}) and (\ref{Hsplit})
(and/or possibly analogous equations for other observables) and
determining model parameters such as $I_1,I_2,\alpha ,\beta, \gamma$
from experimental data.

Taking into account recent experimental observations of the mass of
the $\Theta^{+}$, the parameters entering Eq.(\ref{Hsplit}) can be
conveniently parameterized in terms of the pion-nucleon $\Sigma_{\pi
  N}$ term (assuming $m_{s}/(m_{u}+m_{d})=12.9$) as
\cite{Praszalowicz:2004dn}:
\begin{equation}
\alpha=336.4-12.9\,\Sigma_{\pi N},\quad\beta=-336.4+4.3\,\Sigma_{\pi N}%
,\quad\gamma=-475.94+8.6\,\Sigma_{\pi N}\label{albega}%
\end{equation}
(in units of MeV). Moreover, the inertia parameters which describe the
representation splittings%
\begin{equation}
\Delta M_{{10}-8}=\frac{3}{2I_{1}},\;\;\Delta M_{{\overline{10}}-8}%
=\frac{3}{2I_{2}}%
\end{equation}
take the following values (in MeV):%
\begin{equation}
\frac{1}{I_{1}}=152.4,\quad\frac{1}{I_{2}}=608.7-2.9\,\Sigma_{\pi
N}.\label{ISigma}%
\end{equation}
Equations (\ref{albega}) and (\ref{ISigma}) follow from the fit to the
masses of the octet and decuplet baryons as well as that of the
$\Theta^{+}$.  If, furthermore, one imposes additional constraint that
$M_{\Xi_{\overline{10}}}=1860$ MeV, then $\Sigma_{\pi N}=73$ MeV
\cite{Praszalowicz:2004dn} (see also \cite{Schweitzer:2003fg}) in
agreement with recent experimental estimates~\cite{Sigma}.

Since the symmetry-breaking term (\ref{Hsplit}) of the collective
Hamiltonian mixes different SU(3) representations, the collective wave
functions are given as the following linear combinations
\cite{Kim:1998gt}:
\begin{align}
\left|  B_{8}\right\rangle  &  =\left|  8_{1/2},B\right\rangle +c_{\overline
{10}}^{B}\left|  \overline{10}_{1/2},B\right\rangle +c_{27}^{B}\left|
27_{1/2},B\right\rangle ,\nonumber\\
\left|  B_{10}\right\rangle  &  =\left|  10_{3/2},B\right\rangle +a_{27}%
^{B}\left|  27_{3/2},B\right\rangle +a_{35}^{B}\left|  35_{3/2},B\right\rangle
,\nonumber\\
\left|  B_{\overline{10}}\right\rangle  &  =\left|  \overline{10}%
_{1/2},B\right\rangle +d_{8}^{B}\left|  8_{1/2},B\right\rangle +d_{27}%
^{B}\left|  27_{1/2},B\right\rangle +d_{\overline{35}}^{B}\left|
\overline{35}_{1/2},B\right\rangle, \label{admix}
\end{align}
where $\left|  B_{\mathcal{R}}\right\rangle $ denotes the state which reduces
to the SU(3) representation $\mathcal{R}$ in the formal limit $m_{s}%
\rightarrow0$ and the spin index $J_{3}$ has been suppressed. The
$m_{s}$-dependent (through the linear $m_{s}$ dependence of $\alpha$,
$\beta$ and $\gamma$) coefficients in Eq.(\ref{admix}) read:
\begin{align}
c_{\overline{10}}^{B} &  =c_{\overline{10}}\left[  \kern -0.5em%
\begin{array}
[c]{c}%
\sqrt{5}\\
0\\
\sqrt{5}\\
0
\end{array}
\kern -0.2em\right]  \kern -0.2em,\;c_{27}^{B}=c_{27}\left[  \kern -0.5em%
\begin{array}
[c]{c}%
\sqrt{6}\\
3\\
2\\
\sqrt{6}%
\end{array}
\kern -0.2em\right]  \kern -0.2em,\;a_{27}^{B}=a_{27}\left[  \kern -0.5em%
\begin{array}
[c]{c}%
\sqrt{15/2}\\
2\\
\sqrt{3/2}\\
0
\end{array}
\kern -0.2em\right]  \kern -0.2em,\;a_{35}^{B}=a_{35}\left[  \kern -0.5em%
\begin{array}
[c]{c}%
5/\sqrt{14}\\
2\sqrt{5/7}\\
3\sqrt{5/14}\\
2\sqrt{5/7}%
\end{array}
\kern -0.2em\right] \nonumber\\
d_{8}^{B} &  =d_{8}\left[
\begin{array}
[c]{c}%
0\\
\sqrt{5}\\
\sqrt{5}\\
0
\end{array}
\right]  ,\qquad d_{27}^{B}=d_{27}\left[
\begin{array}
[c]{c}%
0\\
\sqrt{3/10}\\
2/\sqrt{5}\\
\sqrt{3/2}%
\end{array}
\right]  ,\qquad d_{\overline{35}}^{B}=d_{\overline{35}}\left[
\begin{array}
[c]{c}%
1/\sqrt{7}\\
3/(2\sqrt{14)}\\
1/\sqrt{7}\\
\sqrt{5/56}%
\end{array}
\right] \label{mix}%
\end{align}
respectively in the basis $[N,\Lambda,\Sigma,\Xi]$, $[\Delta,\Sigma^{\ast},\Xi^{\ast
},\Omega]$, $\left[  \Theta^{+},N_{\overline{10}},\Sigma_{\overline{10}%
},\Xi_{\overline{10}}\right]$ and analogous states in
$\mathcal{R}=27,35,\overline{35}$, and
\begin{align}
c_{\overline{10}}=-\frac{I_{2}}{15}\left(\alpha+\frac{1}{2}\gamma\right),
&  ~c_{27}=-\frac{I_{2}}{25}\left(  \alpha-\frac{1}{6}\gamma\right)
,\nonumber\\
~~ & \nonumber\\
a_{27}=-\frac{I_{2}}{8}\left(  \alpha+\frac{5}{6}\gamma\right), &
~a_{35}=-\frac{I_{2}}{24}\left(  \alpha-\frac{1}{2}\gamma\right),\nonumber\\
~~ & \nonumber\\
d_{8}=\frac{I_{2}}{15}\left(  \alpha+\frac{1}{2}\gamma\right),~~ &
d_{27}=-\frac{I_{2}}{8}\left(  \alpha-\frac{7}{6}\gamma\right)
,~~~d_{\overline{35}}=-\frac{I_{2}}{4}\left(
\alpha+\frac{1}{6}\gamma\right).\label{mix1}%
\end{align}
%%%%%%%%%%%%%%%%%%%%%%%%%%%%%%%%%%%%%%%%%%%%%%%%%%%%%%%%%%%
\section{Magnetic moments in the chiral quark-soliton model}
\label{magmoms}
%%%%%%%%%%%%%%%%%%%%%%%%%%%%%%%%%%%%%%%%%%%%%%%%%%%%%%%%%%%

The collective operator for the magnetic moments can be parameterized
by six constants By definition in the {\em model-independent approach}
they are treated as free
~\cite{Kim:1997ip,Kim:1998gt}:%
\begin{align}
\hat{\mu}^{(0)} &  =w_{1}D_{Q3}^{(8)}\;+\;w_{2}d_{pq3}D_{Qp}^{(8)}%
\cdot\hat{J}_{q}\;+\;\frac{w_{3}}{\sqrt{3}}D_{Q8}^{(8)}\hat{J}_{3},\\
\hat{\mu}^{(1)} &  =\frac{w_{4}}{\sqrt{3}}d_{pq3}D_{Qp}^{(8)}D_{8q}%
^{(8)}+w_{5}\left(  D_{Q3}^{(8)}D_{88}^{(8)}+D_{Q8}^{(8)}D_{83}^{(8)}\right)
\;+\;w_{6}\left(  D_{Q3}^{(8)}D_{88}^{(8)}-D_{Q8}^{(8)}D_{83}^{(8)}\right)
.\nonumber
\end{align}
The parameters $w_{1,2,3}$ are of order $\mathcal{O}(m_{s}^0)$,
while $w_{4,5,6}$ are of order $\mathcal{O}(m_{s})$, $m_{s}$ being
regarded as a small parameter.

The full expression for the magnetic moments can be decomposed as follows:
\begin{equation}
\mu_{B}=\mu_{B}^{(0)}+\mu_{B}^{(op)}+\mu_{B}^{(wf)},
\end{equation}
where the $\mu_{B}^{(0)}$ is given by the matrix element of the
$\hat{\mu}^{(0)}$ between the purely symmetric states $\left|
\mathcal{R}_{J},B,J_{3}\right\rangle$, and the $\mu_{B}^{(op)}$ is
given as the matrix element of the $\hat{\mu}^{(1)}$ between the
symmetry states as well.  The wave function correction
$\mu_{B}^{(wf)}$ is given as a sum of the interference matrix elements
of the $\mu_{B}^{(0)}$ between purely symmetric states and admixtures
displayed in Eq.(\ref{admix}).  These matrix elements were calculated
for octet and decuplet baryons in Ref.\cite{Kim:1998gt}.  It has been
shown that the $\mu_{B}^{(0)}$ for these two representations depend
only upon the following combinations:
\begin{equation}
v=\frac{1}{60}\left(  w_{1}-\frac{w_{2}}{2}\right)  \quad\text{and}\quad
w=\frac{w_{3}}{120}.\label{vw}%
\end{equation}
Therefore, in the leading order in $m_{s}$, it is impossible to
extract information on $w_{1}$ and $w_{2}$ separately.  In
contrast, the wave function corrections $\mu_{B}^{(wf)}$ depend
separately on all three zeroth-order parameters $w_{1,2,3}$.
However, prior to the discovery of the $\Theta^{+} $, both $I_{2}$
and one of the parameters entering Eq.(\ref{Hsplit}), which we
have chosen to be $\gamma$, were unconstrained, since they did not
enter the formulae for the nonexotic mass splittings. Therefore,
the extraction of $w_{2}$ from the $\mu_{B}^{(wf)}$ was not
possible as well.

In order to make numerical estimates, we have assumed in
Refs.\cite{Kim:1997ip,Kim:1998gt} that $\gamma=0$.  This assumption
was based on the numerical results of the model calculations as well
as on the model value of the $\Sigma_{\pi N}$ being of order of 54 MeV
\cite{Diakonov:1988mg}.  Moreover, in the nonrelativistic limit of the
chiral quark-soliton model $\gamma\equiv0$. This choice reduced the
number of free parameters to seven (six constants $w_{i}$ and
$I_{2}$).  However, due to an accidental algebraical property, the
explicit formulae for the octet magnetic moments depend effectively
only on six parameters. On the contrary, the magnetic moments of the
decuplet depend on all seven parameters and therefore one could
determine them only up to one unknown constant which we called $p$ in
Refs.~\cite{Kim:1997ip,Kim:1998gt}.  Unfortunately, the dependence on
$p$ of the two measured magnetic moments of $\Omega^{-}$ and
$\Delta^{++}$ is too weak to determine $p$.

The situation in the $\overline{10}$ multiplet is very different.
In this case, the $\mu_{B}^{(0)}$ depend on a different combination of
parameters $w_{1}$ and $w_{2}$, hence the prediction for
$\mu_{\Theta^{+}}$ depends on one unknown constant already in the
SU(3)-symmetry limit:
\begin{equation}
\mu_{B}^{{\overline{10}}\;(0)}=\left[  \frac{5}{2}\left(  -v+w\right)
-\frac{1}{8}w_{2}\,\right]  Q_{B}.\label{10bchiral0}%
\end{equation}
Since, as explained above, prior to the measurement of the
$\Theta^{+}$ mass, the determination of $w_{2}$ from the nonexotic
data was not possible, we have assumed in Ref.\cite{Kim:2003ay},
following explicit model calculations~\cite{Kim:1998gt}, that the
parameter $w_{2}$ took the value $w_{2}\simeq5$. This assumption
led to a small but positive value of the magnetic moment of
$\Theta^{+}$.  Surprisingly, we have observed that in the
nonrelativistic limit of the chiral quark-soliton
model~\cite{limit} all three parameters $w_{i}$ can be essentially
expressed in terms of one unknown constant $K$.  This feature
leads to the remarkable result that the magnetic moment of the
positively charged $\Theta^+$ is negative:
$\mu_{\Theta^{+}}^{(0)}<0$.

The measurement of the $\Theta^{+}$ mass constrains the parameter
space of the model in Eq.(\ref{albega}) and Eq.(\ref{ISigma}).
Recent phenomenological analyses indicate that our previous
assumption on $\gamma$, i.e. $\gamma=0$, has to be most likely
abandoned. Therefore, our previous results for the magnetic
moments of $8$, $10$ and $\overline{10}$ have to be reanalyzed. In
the present work we show that a \emph{model-independent} analysis
with this new phenomenological input yields $w_{2}$ much larger
than initially assumed, which causes $\mu_{\Theta^{+}}^{(0)}$ for
realistic values of $\Sigma_{\pi N}$ to be negative and rather
small.  We also show that our previous results for the decuplet
magnetic moments still hold within the accuracy of the model.

The octet and decuplet magnetic moments were calculated in
Refs.\cite{Kim:1997ip,Kim:1998gt}.  For the antidecuplet
$\mu_{B}^{{\overline{10}}\;(0)}$ are given in Eq.(\ref{10bchiral0}).
In order to calculate the $\mu_{B}^{(op)}$, the following relations, 
obtained using  SU(3) Clebsch-Gordan coefficients \cite{KW}, hold:
\begin{align}
D_{33}^{(8)}D_{88}^{(8)}  &  =\frac{1}{5}D_{\Sigma^{0}\Sigma^{0}}%
^{(8)}+\frac{1}{4}D_{\Sigma^{0}\Sigma^{0}}^{(10)}+\frac{1}{4}D_{\Sigma
^{0}\Sigma^{0}}^{(\overline{10})}+\frac{3}{10}D_{\Sigma^{0}\Sigma^{0}}%
^{(27)},\nonumber\label{D3388}\\
D_{38}^{(8)}D_{83}^{(8)}  &  =\frac{1}{5}D_{\Sigma^{0}\Sigma^{0}}%
^{(8)}-\frac{1}{4}D_{\Sigma^{0}\Sigma^{0}}^{(10)}-\frac{1}{4}D_{\Sigma
^{0}\Sigma^{0}}^{(\overline{10})}+\frac{3}{10}D_{\Sigma^{0}\Sigma^{0}}%
^{(27)},\nonumber\\
D_{83}^{(8)}D_{88}^{(8)}  &  =-\frac{1}{5}D_{\Lambda\Sigma^{0}}^{(8)}%
+\frac{9}{20}D_{\Lambda\Sigma^{0}}^{(27)}%
\end{align}
and
\begin{equation}
\frac{1}{\sqrt{3}}d_{ab3}D_{Qa}^{(8)}D_{8b}^{(8)}=\frac{1}{10}\left[
D_{\Sigma^{0}\Sigma^{0}}^{(8)}-D_{\Sigma^{0}\Sigma^{0}}^{(27)}-\frac{1}%
{\sqrt{3}}D_{\Lambda\Sigma^{0}}^{(8)}-\frac{3}{2\sqrt{3}}D_{\Lambda\Sigma^{0}%
}^{(27)}\right]  .
\end{equation}
Furthermore, in order to calculate the $\mu_{B}^{(wf)}$, several
off-diagonal matrix elements of the $\hat{\mu}^{(0)}$ are
required. These have been calculated in
Ref.\cite{Praszalowicz:2004dn} in the context of the hadronic
decay widths of the baryon antidecuplet.

Denoting the set of the model parameters by
\begin{equation}
\vec{w}=(w_{1},\ldots,w_{6})
\end{equation}
the model formulae for the set of the magnetic moments in representation
$\mathcal{R}$ (of dimension $R$)%
\begin{equation}
\vec{\mu}^{\mathcal{R}}=(\mu_{B_{1}},\ldots,\mu_{B_{R}})
\end{equation}
can be conveniently cast into the form of the matrix equations:%
\begin{equation}
\vec{\mu}^{\mathcal{R}}=A^{\mathcal{R}}[\Sigma_{\pi N}]\cdot\vec{w},%
\end{equation}
where rectangular matrices $A^{8}$ and $A^{10}$ can be found in
Refs.\cite{Kim:1997ip,Kim:1998gt}. Note their dependence on the
pion-nucleon $\Sigma_{\pi N}$ term.  As for the antidecuplet, we
find $A^{\overline{10}}$ in the following form:
\begin{equation}
\left[
\begin{array}
[c]{cccccc}%
-\frac{1}{24}+\frac{d_{\overline{35}}}{84} & -\frac{5}{48}-\frac{d_{\overline
{35}}}{168} & \frac{1}{48}+\frac{d_{\overline{35}}}{56} & \frac{1}{56} &
-\frac{1}{84} & 0\\
-\frac{1}{24}-\frac{7\,d_{27}}{72}+\frac{d_{\overline{35}}}{112} &
-\frac{5}{48}+\frac{11\,d_{27}}{144}-\frac{d_{\overline{35}}}{224} &
\frac{1}{48}+\frac{d_{27}}{48}+\frac{3\,d_{\overline{35}}}{224} &
\frac{1}{189} & -\frac{1}{63} & 0\\
-\frac{c_{\overline{10}}}{3}+\frac{7\,d_{27}}{180}+\frac{d_{\overline{35}}%
}{56} & -\frac{c_{\overline{10}}}{3}-\frac{11\,d_{27}}{360}-\frac{d_{\overline
{35}}}{112} & -\frac{c_{\overline{10}}}{6}-\frac{d_{27}}{120}%
+\frac{3\,d_{\overline{35}}}{112} & -\frac{5}{1512} & \frac{13}{252} & 0\\
-\frac{1}{24}-\frac{7\,d_{27}}{36}+\frac{d_{\overline{35}}}{168} &
-\frac{5}{48}+\frac{11\,d_{27}}{72}-\frac{d_{\overline{35}}}{336} &
\frac{1}{48}+\frac{\,d_{27}}{24}+\frac{d_{\overline{35}}}{112} &
-\frac{11}{1512} & -\frac{5}{252} & 0\\
-\frac{c_{\overline{10}}}{6}-\frac{7\,d_{27}}{90}+\frac{d_{\overline{35}}}{84}
& -\frac{c_{\overline{10}}}{6}+\frac{11\,d_{27}}{180}-\frac{d_{\overline{35}}%
}{168} & -\frac{c_{\overline{10}}}{12}+\frac{d_{27}}{60}+\frac{d_{\overline
{35}}}{56} & -\frac{1}{189} & \frac{1}{63} & 0\\
\frac{1}{24}-\frac{c_{\overline{10}}}{3}+\frac{7\,d_{27}}{180}%
+\frac{d_{\overline{35}}}{56} & \frac{5}{48}-\frac{c_{\overline{10}}}%
{3}-\frac{11\,d_{27}}{360}-\frac{d_{\overline{35}}}{112} & -\frac{1}%
{48}-\frac{c_{\overline{10}}}{6}-\frac{\,d_{27}}{120}+\frac{3\,d_{\overline
{35}}}{112} & -\frac{5}{1512} & \frac{13}{252} & 0\\
-\frac{1}{24}-\frac{7\,d_{27}}{24}+\frac{d_{\overline{35}}}{336} &
-\frac{5}{48}+\frac{11\,d_{27}}{48}-\frac{d_{\overline{35}}}{672} &
\frac{1}{48}+\frac{d_{27}}{16}+\frac{\,d_{\overline{35}}}{224} &
-\frac{5}{252} & -\frac{1}{42} & 0\\
-\frac{7\,d_{27}}{36}+\frac{d_{\overline{35}}}{168} & \frac{11\,d_{27}}%
{72}-\frac{d_{\overline{35}}}{336} & \frac{d_{27}}{24}+\frac{d_{\overline{35}%
}}{112} & -\frac{11}{1512} & -\frac{5}{252} & 0\\
\frac{1}{24}-\frac{7\,d_{27}}{72}+\frac{d_{\overline{35}}}{112} & \frac{5}%
{48}+\frac{11\,d_{27}}{144}-\frac{d_{\overline{35}}}{224} & -\frac{1}%
{48}+\frac{\,d_{27}}{48}+\frac{3\,d_{\overline{35}}}{224} & \frac{1}{189} &
-\frac{1}{63} & 0\\
\frac{1}{12}+\frac{d_{\overline{35}}}{84} & \frac{5}{24}-\frac{d_{\overline
{35}}}{168} & -\frac{1}{24}+\frac{d_{\overline{35}}}{56} & \frac{1}{56} &
-\frac{1}{84} & 0
\end{array}
\right]
\label{Aa10}
\end{equation}
in the basis
\begin{equation}
\vec{\mu}^{\overline{10}}=(\mu_{\Theta^{+}},\mu_{p^{\ast}},\mu_{n^{\ast}}%
,\mu_{\Sigma^{+}},\mu_{\Sigma^{0}},\mu_{\Sigma^{-}},\mu_{\Xi^{+}},\mu_{\Xi
^{0}},\mu_{\Xi^{-}},\mu_{\Xi^{--}}).
\end{equation}

%%%%%%%%%%%%%%%%%%%%%%%%%%%%%%%%%%%%%%%%%%%%%%
\section{Results and discussion}\label{numres}
%%%%%%%%%%%%%%%%%%%%%%%%%%%%%%%%%%%%%%%%%%%%%%
In order to find the set of parameters $w_{i}[\Sigma_{\pi N}]$, we
minimize the mean square deviation for the octet magnetic moments:
\begin{equation}
\Delta\mu^{8}=\frac{1}{7}\sqrt{\sum_{B}\left(
    \mu_{B,\,th}^{8}[\Sigma_{\pi N}]-\mu_{B,\,exp}^{8}\right)  ^{2}},
\end{equation}
where the sum extends over all octet magnetic moments, but the
$\Sigma^{0}$.  The value $\Delta\mu^{8}\simeq0.01$ is in practice
not sensitive to the $\Sigma_{\pi N}$ in the physically interesting range
$45$ $-$ $75$ MeV.  Therefore, the values of the
$\mu_{B,\,th}^{8}[\Sigma_{\pi N}]$ are also not sensitive to
$\Sigma_{\pi N}$.  Table~\ref{tab:1} lists the results of the magnetic
moments of the baryon octet. 
\begin{table}[h]
  \centering
  \begin{tabular}[c]{|l|ccccccc|}\hline
& $p$ & $n$ & $\Lambda^{0}$ & $\Sigma^{+}$ & $\Sigma^{-}$ & $\Xi^{0}$
& $\Xi^{-}$ \\ \hline
th. & $2.814$ & $-1.901$ & $-0.592$ & $2.419$ & $-1.172$ & $-1.291$ & $-0.656$\\
exp. & $2.793$ & $-1.913$ & $-0.613$ & $2.458$ & $-1.16$~ &
$-1.25$~ & $-0.651$ \\ \hline
  \end{tabular}
  \caption{Magnetic moments of the baryon octet.}
\label{tab:1}
\end{table}

Similarly, the value of the nucleon strange magnetic moment is not
sensitive to $\Sigma_{\pi N}$ and reads $\mu_{N}^{(s)}=0.39 \,{\rm
  n.m.}$ in fair agreement with our previous analysis of
Ref.\cite{Kim:1998gt}.  Parameters $w_{i}$, however, do depend on
$\Sigma_{\pi N}$.  This is 
shown in Table.\ref{tablewi}:
\begin{table}[h]
\begin{tabular}
[c]{|c|cccccc|}\hline
\multicolumn{1}{|c|}{$\Sigma_{\pi N}$ [MeV]} & $w_{1}$ & $w_{2}$ &
$w_{3}$ &   $w_{4}$ & $w_{5}$ & $w_{6}$ \\ \hline
$45$ & $-8.564$ & $14.983$ & $7.574$ & $-10.024$ & $-3.742$ & $-2.443$\\
$60$ & $-10.174$ & $11.764$ & $7.574$ & $-9.359$ & $-3.742$ & $-2.443$\\
$75$ & $-11.783$ & $8.545$ & $7.574$ & $-6.440$ &  $-3.742$ & $-2.443$
\\ \hline
\end{tabular}
\caption{Dependence of the parameters $w_i$ on $\Sigma_{\pi N}$.}
\label{tablewi}
\end{table}
Note that parameters $w_{2,3}$ are formally
$\mathcal{O}(1/N_{c})$ with respect to $w_{1}$. For smaller
$\Sigma_{\pi N}$, this $N_{c}$ counting is not borne by explicit
fits. Interestingly, the chiral-limit parameters $v$ and $w$
defined in Eq.(\ref{vw}) do not depend on $\Sigma_{\pi N}$ and
read:
\begin{equation}
v=-0.268,\quad w=0.063.\label{vwnum}%
\end{equation}
The values of $v$ and $w$ in Eq.(\ref{vwnum}) almost exactly coincide with the
parameters extracted from the linear combinations%
\begin{equation}%
\begin{array}
[c]{ccrcc}%
v & = & \left(  2\mu_{\mathrm{n}}-\mu_{\mathrm{p}}+3\mu_{\Xi^{0}}+\mu_{\Xi
^{-}}-2\mu_{\Sigma^{-}}-3\mu_{\Sigma^{+}}\right)  /60 & = & -0.268,\\
w & = & \left(  3\mu_{\mathrm{p}}+4\mu_{\mathrm{n}}+\mu_{\Xi^{0}}-3\mu
_{\Xi^{-}}-4\mu_{\Sigma^{-}}-\mu_{\Sigma^{+}}\right)  /60 & = & 0.060.
\end{array}
\label{Eq:mean}%
\end{equation}
which are free of linear $m_{s}$ corrections~\cite{Kim:1998gt}.
This is a remarkable feature of
the present fit, since when the $m_{s}$ corrections are included, the
$m_{s}$-independent parameters need not be refitted.  This property
will be used in the following when we restore the linear dependence of
the $\mu_{B}^{\overline{10}}$ on $m_{s}$.

\qquad\ The magnetic moments of the baryon decuplet and
antidecuplet depend on the $\Sigma_{\pi N}$.  However, the
dependence of the decuplet is very weak.  The results are
summarized in Table \ref{table10},
\begin{table}[h]
\begin{tabular}
[c]{|c|cccccccccc|}\hline
$\Sigma_{\pi N}$ [MeV] & $\Delta^{++}$ & $\Delta^{+}$ & $\Delta^{0}$ &
$\Delta^{-}$ & $\Sigma^{\ast+}$ &
$\Sigma^{\ast0}$ & $\Sigma^{\ast-}$ & $\Xi^{\ast0}$ & $\Xi^{\ast-}$ &
$\Omega^{-}$\\\hline
$45$ & $5.40$ & $2.65$ & $-0.09$ & $-2.83$ & $2.82$ & $0.13$ & $-2.57$
& $0.34$ & $-2.31$ & $-2.05$\\
$60$ & $5.39$ & $2.66$ & $-0.08$ & $-2.82$ & $2.82$ & $0.13$ & $-2.56$
& $0.34$ & $-2.30$ & $-2.05$\\
$75$ & $5.39$ & $2.66$ & $-0.07$ & $-2.80$ & $2.81$ & $0.13$ & $-2.55$
& $0.33$ & $-2.30$ & $-2.05$\\
\text{Ref.\cite{Kim:1997ip}} & $5.34$ & $2.67$ & $-0.01$ & $-2.68$ &
$3.10$ & $0.32$ & $-2.47$ & $0.64$ & $-2.25$ & $-2.04$ \\ \hline
\end{tabular}
\caption{Magnetic moments of the baryon decuplet.}
\label{table10}%
\end{table}
where we also display the theoretical predictions from
Ref.\cite{{Kim:1997ip}} for $p=0.25$.  Let us note that the $m_{s}$
corrections are not large for the decuplet and the approximate
proportionality of the $\mu_{B}^{10}$ to the baryon charge
$Q_{B}$ still holds.

\begin{figure}[h]
\begin{center}
\includegraphics[scale=1.1]{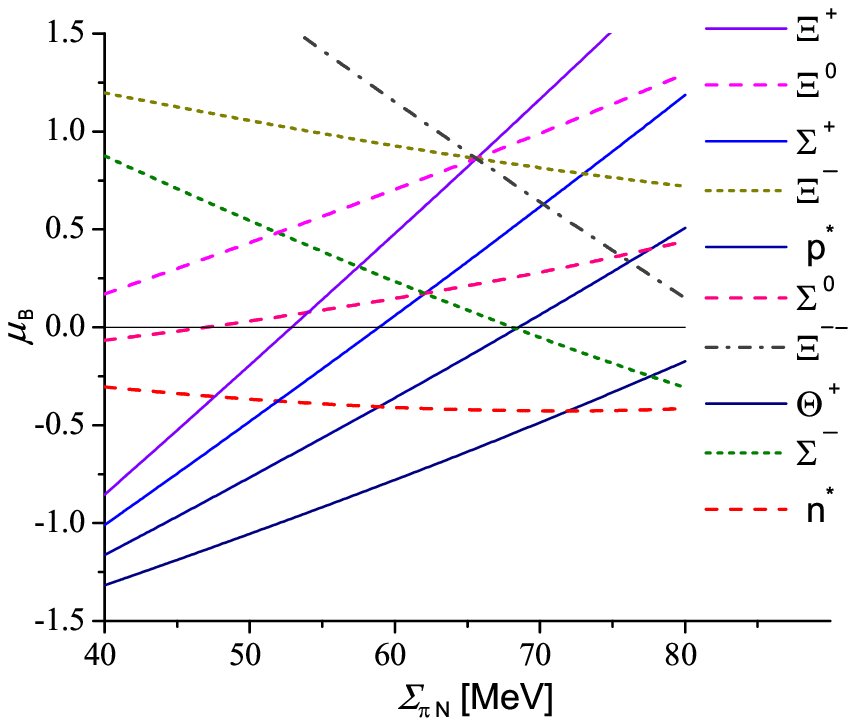}
\end{center}
\caption{Magnetic moments of antidecuplet as functions of $\Sigma_{\pi N}$.}%
\label{fig:10bar}%
\end{figure}

Finally,  for antidecuplet we have a strong dependence on
$\Sigma_{\pi N}$, yielding the numbers of Table \ref{tab10b}.
\begin{table}[h]
\begin{tabular}
[c]{|c|cccccccccc|}\hline
$\Sigma_{\pi N}$ [MeV] & $\Theta^{+}$ & $p^{\ast}$ & $n^{\ast}$ &
$\Sigma_{\overline{10}}^{+}$ & $\Sigma_{\overline{10}}^{0}$ &
$\Sigma_{\overline{10}}^{-}$ & $\Xi_{\overline{10}}^{+}$ &
$\Xi_{\overline{10}}^{0}$ & $\Xi_{\overline{10}}^{-}$ &
$\Xi_{\overline {10}}^{--}$ \\\hline
$45$ & $-1.19$ & $-0.97$ & $-0.34$ & $-0.75$ & $-0.02$ & $\;0.71$ &
$-0.53$ & $0.30$ & $1.13$ & $1.95$\\
$60$ & $-0.78$ & $-0.36$ & $-0.41$ & $\;0.06$ & $\;0.15$ & $\;0.23$ &
$\;0.48$ & $0.70$ & $0.93$ & $1.15$\\
$75$ & $-0.33$ & $\;0.28$ & $-0.43$ & $\;0.90$ & $\;0.36$ & $-0.19$ &
$\;1.51$ & $1.14$ & $0.77$ & $0.39$ \\ \hline
\end{tabular}
\caption{Magnetic moments of the baryon antidecuplet.}
\label{tab10b}%
\end{table}
The results listed in Table~\ref{tab10b} are further depicted in
Fig.\ref{fig:10bar}.

%%%%%%%%%%%%%%%%%%%%%%%%%%%%%%%%%%%%%%%%%%%%%%%%%%%%%%%%%
%\vspace{0.5cm}
%%%%%%%%%%%%%%%%%%%%%%%%%%%%%%%%%%%%%%%%%%%%%%%%%%%%%%%%%

\begin{figure}[h]
\begin{center}
\includegraphics[scale=1]{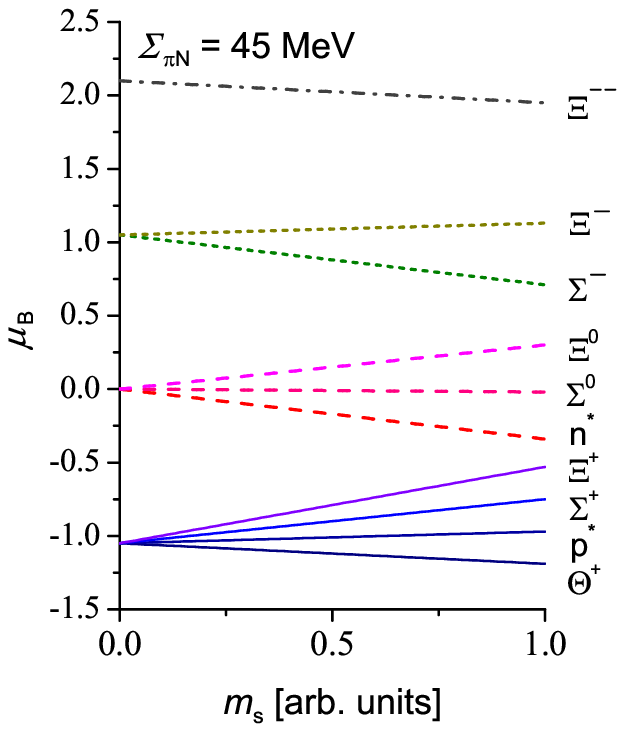}
\includegraphics[scale=1]{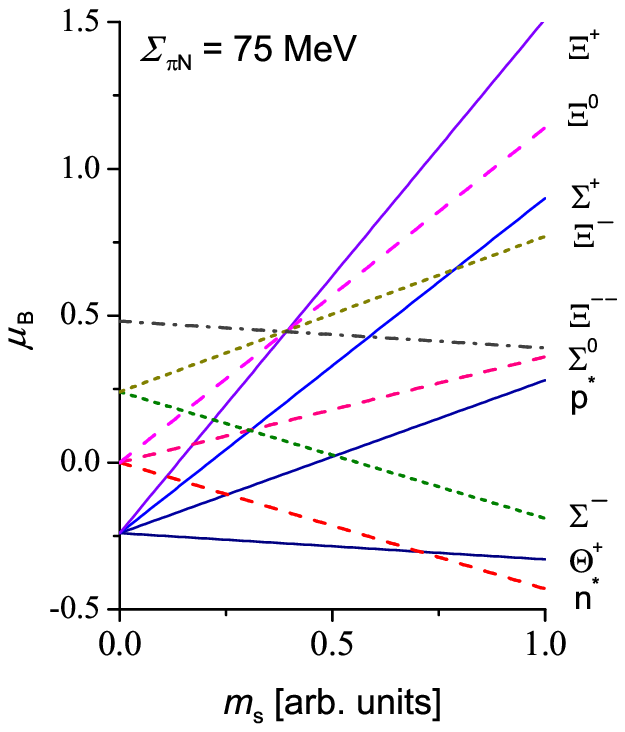}
\end{center}
\caption{Dependence of magnetic moments on $m_{s}$ for $\Sigma_{\pi N}=45$ and
$75$ MeV.}%
\label{fig:10bms}%
\end{figure}

In the chiral limit, the antidecuplet magnetic moments are
proportional to the corresponding charges, see
Eq.(\ref{10bchiral0}), but with opposite sign, and they read
numerically
\begin{equation}
\mu_{B}^{{\overline{10}}\;(0)}=-(1.05\sim 0.24)Q_{B}\label{10bchiral}%
\end{equation}
for $\Sigma_{\pi N}=45$ and $75$ MeV, respectively. The inclusion
of the $m_{s}$ corrections introduces splittings and
proportionality to the charge is violated.  The magnitude of the
splittings increases with $\Sigma_{\pi N}$.  This is depicted in
Fig.\ref{fig:10bms}, where linear dependence on $m_{s}$ is
reproduced from the knowledge of two points:
$\mu_{B}^{\overline{10}}$ in the chiral limit for $m_{s}=0$
(\ref{10bchiral}) and for physical $m_{s}=1$ in arbitrary units as
given in Table \ref{tab10b}.  We see that for small $\Sigma_{\pi
N}$ corrections due to the nonzero $m_{s}$ are moderate and the
perturbative approach is reliable.  On the contrary, for large
$\Sigma_{\pi N}$, corrections are large. This is due to the wave
function corrections, since the dependence of the operator part on
the $\Sigma_{\pi N}$ given in terms of the coefficients
$w_{4,5,6}$ is small as in Table \ref{tablewi}.  The wave function
corrections cancel for the non-exotic baryons and add
constructively for the baryon antidecuplet.  In particular, for
$\Sigma_{\pi N}=75$~MeV we have large admixture coefficient of
27-plet: $d_{27}^{B}$ tends to dominate otherwise small magnetic
moments of antidecuplet.  At this point, the reliability of the
perturbative expansion for the antidecuplet magnetic moments may
be questioned.  On the other hand, as remarked above, the $N_{c}$
counting for the $w_{i}$ coefficients works much better for large
$\Sigma_{\pi N}$.  One notices for reasonable values of
$\Sigma_{\pi N}$ some interesting facts, which were partially
reported already in Ref.\cite{Kim:2003ay}: The magnetic moments of
the antidecuplet baryons are rather small in absolute value. For
$\Theta^+$ and $p^*$ one obtains negative values although the
charges are positive.  For $\Xi^{-}_{\overline{10}}$ and
$\Xi^{--}_{\overline{10}}$ one obtains positive values although
the signs of the charges are negative.

%%%%%%%%%%%%%%%%%%%%%%%%%%%%%%%%%%%%%%%%%%%%%%%%%%%%%%%%%
%\vspace{0.5cm}
%%%%%%%%%%%%%%%%%%%%%%%%%%%%%%%%%%%%%%%%%%%%%%%%%%%%%%%%%
\section{Conclusion and summary}\label{summary}
%%%%%%%%%%%%%%%%%%%%%%%%%%%%%%%%%%%%%%%%%%%%%%%%%%%%%%%%%

The magnetic moments of the positive-parity pentaquarks have been
studied by a number of authors in different models
\cite{Zhao:2003gs,Huang:2003bu,Liu:2003ab,Bijker:2004gr,Hong:2004xn}.
The results are displayed in Table~\ref{table5}.  We see that in
all quark models the magnetic moment of the $\Theta^+$ is rather
small and positive.  On the contrary, our present analysis shows
that $\mu_{\Theta^+}< 0$, although the magnitude depends strongly
on the model parameters.  The measurement of $\mu_{\Theta^+}$
could therefore discriminate between different models. This also
may add to reduce the ambiguities in the pion-nucleon sigma term
$\Sigma_{\pi N}$.

The measurement of the antidecuplet magnetic moments by ordinary
precession techniques is not possible.  However, it is crucial to
know the magnetic moment of the $\Theta^{+}$ in order to study its
production via photo-reactions. One can use the measured cross
section to determine the magnetic moment of the $\Theta^{+}$.  The
cross sections for the $\Theta^{+}$ production from nucleons
induced by photons~\cite{Nametal} have been already described
theoretically.  A similar approach was used to determine the
magnetic moments of the $\Delta^{++}$~\cite{Deltapp1,Deltapp2} and
$\Delta^{+}$~\cite{Kotulla:2002cg}, which are much broader than
the $\Theta^{+}$.  The measurements of the $\Delta^{++}$ magnetic
moment comes from the reaction such as
$\pi^{+}p\rightarrow\pi^{+}\gamma^{\prime}p$~\cite{Deltapp1,Deltapp1},
while that of the $\Delta^{+}$ was measured in $\gamma
p\rightarrow\pi^{0}\gamma^{\prime}p$~\cite{Kotulla:2002cg}. This
shows that the measurement of the magnetic moments of resonances
is in principle possible, despite the fact that it is difficult
and is hampered by large uncertainties which mainly come from the
systematic error of cross-section calculations.

\begin{table}
\label{table5}
\begin{tabular}[c]{|cl|lcccc|}\hline
Ref. & first author & model & remarks & ~$\Theta^{+}$ & ~$\Xi^{--}$ & ~$\Xi^{+}%
$\\ \hline
\cite{Zhao:2003gs} & Q. Zhao & diquarks (JW) &  &  $~~0.08$ &  $-$ &  $-$\\
\hline
\cite{Huang:2003bu}& P.Z. Huang & sum rules & abs. value &  $0.12\pm0.06$ &  $-$ &  $-$\\
\hline
\cite{Liu:2003ab}& Y.-R. Liu & diquarks (JW) &  &  $~~0.08$ &  $~~0.12$ &  $-0.06$\\
&  & clusters (SZ) &  &  $~~0.23$ &  $-0.17$ &  $~~0.33$\\
&  & triquarks (KL) &  &  $~~0.37$ &  $~~0.43$ &  $~~0.13$\\
&  & MIT bag (S) &  &  $~~0.37$ &  $-0.42$ &  $~~0.45$\\
\hline
\cite{Bijker:2004gr}& R. Bijker & QM, harm.osc. &  &  $~~0.38$ &  $-0.44$ &  $~~0.50$\\
\hline
\cite{Hong:2004xn}& D.K. Hong & chiral eff. th. &  $m_{s}=400$ MeV &  $~~0.71$ &  $-$ &  $-$\\
&  & with diquarks &  $m_{s}=450$ MeV &  $~~0.56$ &  $-$ &  $-$\\
\hline
\multicolumn{2}{|c|}{present  work}& chiral soliton &  $\Sigma_{\pi N}=45$ MeV &  $-1.19$ &
$~~1.95$ &  $-0.53$\\
&  & model & $\Sigma_{\pi N}=75$ MeV &  $-0.33$ &  $~~0.39$ &
$~~1.51$ \\ \hline
\end{tabular}
\caption{Magnetic moments of $\Theta^+$, $\Xi^{--}$ and $\Xi^{+}$ in nuclear
magnetons from different papers in different models. (JW) stands for Jaffe
and Wilczek \cite{Jaffe:2003sg}, (SZ) for Shuryak and Zahed \cite{Shuryak:2003zi},
(KL) for Karliner and Lipkin \cite{Karliner:2003dt} and (S) for Strottman
\cite{Strottman:1979qu}.}
\end{table}

In the present work, we determined the magnetic moments of the
baryon antidecuplet in the \emph{model-independent} analysis
within the chiral quark-soliton model, i.e. using the rigid-rotor
quantization with the linear $m_{s}$ corrections included.
Starting from the collective operators with dynamical parameters
fixed by experimental data, we obtained the magnetic moments of
the baryon antidecuplet (\ref{Aa10}).  The expression for the
magnetic moments of the baryon antidecuplet is different from
those of the baryon decuplet. We found that the magnetic moment
$\mu_{\Theta^{+}}$ is negative and rather strongly dependent on
the value of the $\Sigma_{\pi N}$.  Indeed, the $\mu
_{\Theta^{+}}$ ranges from $-1.19\,{\rm n.m.}$ to $-0.33\,{\rm
n.m.}$ for $\Sigma_{\pi N} = 45$ and $75$ MeV, respectively.  This
is in contrast with our previous estimate of the
$\mu_{\Theta^{+}}$ in the chiral limit~\cite{Kim:2003ay}, where we
have used $w_{2}\sim5$ motivated by the explicit model
calculations. Indeed, Eq.(\ref{10bchiral0}) yields in this case
$\mu_{B}^{_{\overline{10}}} \sim0.20\,Q_{B}$.

One should note that the magnetic moments of the decuplet do not
differ from our previous estimates \cite{Kim:1997ip}.

\section*{Acknowledgments}
The authors acknowledge the hospitality of the Research Center for Nuclear
Physics (RCNP) at the Osaka University, M.P. acknowledges the hospitality of
the Nuclear Theory Group at Brookhaven National Laboratoty (BNL) where parts
of this work have been completed.

H.-Ch.K is grateful to J.K. Ahn (LEPS collaboration), A. Hosaka,
S.I. Nam, M.V. Polyakov and I.K. Yoo (NA49 collaboration) for
valuable discussions.  The present work is supported by Korea
Research Foundation Grant: KRF-2003-041-C20067 (H.-Ch.K.) and by
the Polish State Committee for Scientific Research under grant 2
P03B 043 24 (M.P.) and by Korean-German and Polish-German grants
of the Deutsche Forschungsgemeinschaft.

\end{document}